# A SPHERICAL HARMONIC APPROACH TO REDSHIFT DISTORTION: IMPLICATIONS FOR $\Omega_\circ$ AND THE POWER SPECTRUM

K. B. FISHER

*Institute of Astronomy, Madingley Rd., Cambridge, CB3 0HA.*

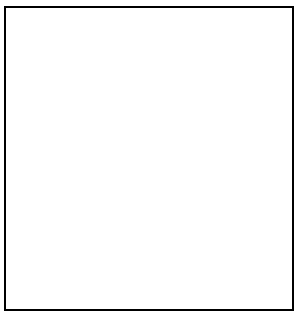

**Abstract**

We examine the nature of galaxy clustering in redshift space using a method based on an expansion of the galaxian density field in Spherical Harmonics and linear theory. We derive a compact and self-consistent expression for the distortion when applied to flux limited redshift surveys. The amplitude of the distortion is controlled by the combination of the density and bias parameters, $\beta \equiv \Omega_\circ^{0.6}/b$ as well as the shape of the real space power spectrum, $P(k)$ (characterized by a shape parameter $\Gamma$), and its normalization, $\sigma_8$; we exploit this fact to derive a maximum likelihood estimator for $\beta$, $\Gamma$, and $\sigma_8$. We check our formalism using $N$-body simulations and demonstrate it provides an unbiased estimate of $\beta$ when the amplitude and shape of the galaxy power spectrum is known. Application of the technique to the 1.2 Jy *IRAS* redshift survey yields $\beta = 0.94 \pm 0.17$ and $\Gamma = 0.17 \pm 0.05$ (1-$\sigma$) when $\sigma_8$ is held fixed at its best value as determined from the real space correlation function. Allowing $\sigma_8$ to be a free parameter, we find $\beta = 0.47 \pm 0.25$, $\Gamma = 0.15 \pm 0.05$, and $\sigma_8 = 0.81 \pm 0.06$.

## 1 Introduction

The clustering of galaxies in redshift space appears systematically different from the clustering that one would observe in real space. On large scales, the coherent motions of galaxies tend to enhance structures tangential to the observer's line-of-sight; conversely, on small scales the peculiar velocities smear out structures along the line of sight giving rise to so-called "Fingers-of-God." Consequently, galaxy clustering is inherently *anisotropic* in redshift space. In the context of linear theory, Kaiser [15] was able derive a simple expression for the angular dependence of the distortion in terms of the Fourier coefficients of the density field,

$$\delta_S(\mathbf{k}) = \delta_R(\mathbf{k}) \left(1 + \beta \mu_{KL}^2\right) ,  \qquad (1)$$

where $\mu_{KL}$ is the cosine of the angle between the wave vector $\mathbf{k}$ and the observer's line-of-sight. In Equation (1) and in the remainder of this paper, we adopt the notation $\beta \equiv \Omega_\circ^{0.6}/b$, where $\Omega_\circ$ and $b$ are the current density and bias (assumed here and below to be independent of scale) parameters respectively; we also will use $R$ and $S$ super and subscripts when referring to quantities in real and redshift space respectively.

The quantity $\beta$ naturally appears in Equation (1) since it is the controls the amplitude of the peculiar velocity field; in linear theory, $\beta$ is the proportionality factor relating the velocity and density fields with $\nabla \cdot \mathbf{v} = -H_\circ \beta \delta$. The dependence of the clustering distortion on the cosmic density parameter has lead to a variety of methods (e.g., [14], [2] and references therein) aimed at exploiting the measurements of the redshift distortion as a way to determine $\Omega_\circ$. Here we present a technique of investigating redshift distortions based on a linear expansion of the galaxian density field in spherical harmonics. The radial nature of the peculiar velocity field leads to a simple and compact formulation of the redshift distortion in spherical coordinates. The work presented here is an extension of the method developed in Fisher, Scharf, & Lahav [11]. Spherical harmonic analyses of the density field (hereafter SHA) have been shown to be a successful way to quantify large scale structure and have been used in a variety of cosmographic applications; an overview of past and present work in the subject can be found in O. Lahav's contribution to these proceedings.

## 2 Weighted Harmonics and Redshift Distortion

We start by defining a weighted spherical harmonic decomposition of the flux-limited density field, in redshift space as,

$$a_{lm}^S = \sum_{i=1}^{N_g} f(\mathbf{s}_i) Y_{lm}(\hat{\mathbf{s}}_i) ,\qquad(2)$$

where $N_g$ is the number of galaxies in the survey, $Y_{lm}$ is the usual spherical harmonic, and $f(s)$ is an arbitrary radial weighting function.[1] In the special case of constant weighting, one simply obtains the projected angular distribution on the sky which is undistorted when the summation in Equation (2) is carried out in redshift space; however, in the more general case where $f(s)$ varies with $s$, peculiar velocities will distort the weighted harmonics.

In order to relate the redshift harmonics given by Equation (2) to their real space counterparts, we first rewrite the summation in Equation (2) as a continuous integral over the density fluctuations in redshift space,

$$a_{lm}^S = \int d^3\mathbf{s}\, \phi(r) f(s) [1 + \delta_S(\mathbf{s})] Y_{lm}(\hat{\mathbf{s}}) .\qquad(3)$$

In Equation (3), the selection function is assumed to be normalized such that $\int dr\, r^2 \phi(r) = N_g/\omega$ where $\omega$ is the solid angle subtended by the survey. Notice that the selection function in Equation (3) is evaluated at the galaxy's *distance*, not redshift, because if the catalogue is flux limited the probability of a galaxy being at redshift, $\mathbf{s}$, will be proportional to the selection function evaluated at the galaxy's actual (albeit unknown) distance, i.e. $\propto \phi(r)$. Next, we note that, by construction, $n_S(\mathbf{s}) d^3\mathbf{s} = n_R(\mathbf{r}) d^3\mathbf{r}$, where $n_S(\mathbf{s})$ and $n_R(\mathbf{r})$ refer to the densities in redshift and real space and that if the perturbations induced by peculiar motions are small, then we can perform a Taylor series expansion of all redshift quantities to first order in the density fluctuation, e.g., $f(\mathbf{s}) \simeq f(r) + \frac{df(r)}{dr}(U(\mathbf{r}) - \mathbf{V}_{obs} \cdot \hat{\mathbf{r}})$, where we follow the notation of Kaiser [15] and define $U(\mathbf{r})$ to be the radial component of the peculiar velocity field, i.e., $U(\mathbf{r}) \equiv \mathbf{v}(\mathbf{r}) \cdot \hat{\mathbf{r}}$; $\mathbf{V}_{obs}$ is the peculiar velocity of the observer at $\mathbf{r} = 0$ (we adopt units where $H_\circ = 1$).

Thus, expansion of Equation (3) yields,

$$\begin{aligned} a_{lm}^S &= \int d^3\mathbf{r}\, \phi(r) f(r) [1 + \delta_R(\mathbf{r})] Y_{lm}(\hat{\mathbf{r}}) \\ &+ \int d^3\mathbf{r}\, \phi(r) \frac{df}{dr} (U(\mathbf{r}) - \mathbf{V}_{obs} \cdot \hat{\mathbf{r}}) Y_{lm}(\hat{\mathbf{r}}) ,\end{aligned}\qquad(4)$$

---

[1] In our analysis $f(s)$ is required to be continuous in its first derivative and to vanish at infinity; this simplifies the analysis by eliminating surface terms that arise when $f(s)$ has a discontinuous boundary.

There are several comments to be made regarding the validity of Equation (4). First, we have assumed that the summation in Equation (2) is carried out over *all* galaxies in a flux limited redshift survey. In this case, the integrals in Equation (4) extend over all space and there are no "surface" terms arising from the deformation of the boundary of the integration region that occurs in the transformation from redshift to real space. Second, there is an apparent absence of terms involving the derivatives of the selection function; these terms would be manifest if we, like Kaiser [15], took $f(s)$ to be the special case of $1/\phi(s)$.

The expansion in Equation (4) can be simplified by expanding the radial peculiar velocity field dependence, $U(\mathbf{r})$, in spherical harmonics:

$$U(\mathbf{r}) = \frac{\beta}{2\pi^2} \sum_{lm} (i^l)^* \int d^3\mathbf{k}\, \frac{\delta^R_\mathbf{k}}{k} j'_l(kr) Y_{lm}(\hat{\mathbf{k}}) Y^*_{lm}(\hat{\mathbf{r}}) \;. \tag{5}$$

In Equation (5) and below, $j'_l(kr) = dj_l(kr)/d(kr)$ refers to the first derivative of the Bessel function. Equation (5) can be derived in a straightforward way by using the linearized continuity equation to relate the velocity to the density field, $\mathbf{v_k} = -i\beta\delta^R_\mathbf{k}\mathbf{k}/k^2$, and the Rayleigh expansion of a plane wave in spherical waves. Using Equation (5), Equation (4) can be written (for $l \geq 1$) as

$$a^S_{lm} = \frac{(i^l)^*}{2\pi^2} \int d^3\mathbf{k}\, \delta^R_\mathbf{k} \left[\Psi^R_l(k) + \beta \Psi^C_l(k)\right] Y_{lm}(\hat{\mathbf{k}}) \;, \tag{6}$$

where

$$\Psi^R_l(k) = \int_0^\infty dr\, r^2 \phi(r) f(r) j_l(kr) \tag{7}$$

describes the real space contribution to the harmonics and

$$\Psi^C_l(k) = \frac{1}{k} \int_0^\infty dr\, r^2 \phi(r) \frac{df(r)}{dr} \left(j'_l(kr) - \frac{1}{3}\delta_{l1}\right) \;, \tag{8}$$

is a "correction" term which embodies the redshift distortions. In Equation (8), $\delta_{l1}$ is a Kronecker delta which contributes only to the dipole ($l = 1$) harmonic. In the case of equal weight, $df/dr = 0$, $\Psi^C_l(k)$ vanishes and Equation (6) leads to the real space expression for the harmonics derived by Scharf *et al.* [22]. The $1/k$ scaling of the distortion window, $\Psi^C_l(k)$ shows that it will generally be much more sensitive to long wavelength perturbations than the real window $\Psi^R_l(k)$.

The expected linear theory *rms* value of the harmonics is given by ensemble average of the square of Equation (6). Using the statistical independence of Fourier waves[2], $\langle \delta^R_\mathbf{k} \delta^{R*}_{\mathbf{k}'}\rangle = (2\pi)^3 P_R(k) \delta^{(3)}(\mathbf{k} - \mathbf{k}')$, this yields

$$\langle |a^S_{lm}|^2 \rangle = \frac{2}{\pi} \int_0^\infty dk\, k^2 P_R(k) \left|\Psi^R_l(k) + \beta \Psi^C_l(k)\right|^2 \;. \tag{9}$$

In real data the square of harmonics in Equation (9) will have a discreteness or "shot" noise contribution; this can be modeled by adding

$$\langle |a_{lm}|^2 \rangle_{SN} = \int_0^\infty dr\, r^2 \phi(r) \left[f(r)\right]^2 \tag{10}$$

to Equation (9).

The $l = 1$ or dipole distortion has two contributions: one from the external dipole moment of the velocity field around the observer and one induced by the motion of observer itself. The latter distortion is caused by $\mathbf{V}_{obs} \neq \mathbf{0}$ and is the origin of the so-called "rocket effect" (c.f., [15], [17], [26]).

---

[2]This assumes that the survey contains many independent modes of the wavenumber in question and usually phrased somewhat loosely as the "fair" sample hypothesis.

We stress, however, that in linear theory the choice of reference frame (i.e., the value of $\mathbf{V}_{obs}$) affects only the value of the dipole harmonic and that for $l \geq 2$ the results are independent of the assumed motion of the observer. The *rms* averaging procedure in Equation (9) treats the observer as a typical point in space and therefore assumes that the observer's peculiar motion is not unusually high or low. This point should be kept in mind if the motion of the Local Group is atypical. In practice, these complications can be avoided by working in the comoving frame of the observer, i.e., the reference frame in which $\mathbf{V}_{obs}$ vanishes or omitting the dipole harmonic in the analysis.

The predicted *rms* redshift harmonics depend on three quantities: the value of $\beta$ and the shape and normalization of the power spectrum. The shape of the power spectrum can be conveniently parametrized by a series of phenomological CDM models with with varying $\Gamma \equiv \Omega_o h$ (e.g., [5]). For the normalization, we adopt the standard $\sigma_8$ convention corresponding to the variance in spheres of 8 $h^{-1}$Mpc.

## 3 Likelihood Estimator

Suppose one computes a set of harmonics, $a^i_{lm}$, for a set of different weighting functions, $f^{(i)}(r)$, for $\{i = 1, \ldots N\}$. If the underlying density field is Gaussian, then the real and imaginary parts of the spherical harmonic coefficients are independent Gaussian variables. In this case, the likelihood (defined as the negative logarithm of the joint probability) is given by

$$\ln \mathcal{L}[\beta, \sigma_8, \Gamma] \equiv -\left(\frac{2l+1}{2}\right) \ln |\det A_{ij}| - \frac{1}{2} \sum_{i,j} A_{ij}^{-1} \left[Re(a^i_{l0}) Re(a^j_{l0})\right]$$

$$- \sum_{i,j} A_{ij}^{-1} \sum_{m=1}^{l} \left[Re(a^i_{lm}) Re(a^j_{lm}) + Im(a^i_{lm}) Im(a^j_{lm})\right] \quad , \tag{11}$$

where $A_{ij}$ is the covariance matrix

$$A_{ij} = \langle a^i_{lm} a^{j*}_{lm} \rangle_{TH} + \langle a^i_{lm} a^{j*}_{lm} \rangle_{SN} \quad , \tag{12}$$

with

$$\langle a^i_{lm} a^{j*}_{lm} \rangle_{TH} = \frac{2}{\pi} \int_0^\infty dk \, k^2 P_R(k) \left(\Psi_l^{R(i)}(k) + \beta \Psi_l^{C(i)}(k)\right) \times$$

$$\left(\Psi_l^{R(j)}(k) + \beta \Psi_l^{C(j)}(k)\right)^* \tag{13}$$

and

$$\langle a^i_{lm} a^{j*}_{lm} \rangle_{SN} = \int_0^\infty dr \, r^2 \phi(r) \left[f^{(i)}(r) f^{(j)}(r)\right] \quad , \tag{14}$$

In Equation (11) $Re(a_{lm})$ and $Im(a_{lm})$ refer to the real and imaginary parts of $a_{lm}$ [23]. The superscripts on $\Psi^R(k)$ and $\Psi^C(k)$ refer to the window function of the corresponding weighting function. Since the harmonics at different $l$ are independent, the total likelihood for $l \leq l_{max}$ is just given by the sum of the likelihoods in Equation (11) for each $l$ up to $l_{max}$. Even if the primordial density is a true Gaussian field, nonlinear evolution will lead to departures from Gaussianity on small scales. Consequently, we expect the formulation of the likelihood given above to be correct (under the assumption of Gaussian primordial fluctuations) only for low order harmonics which probe fluctuations still in the linear regime.

The number and functional form of the weighting functions, $f^{(i)}(r)$, used to construct the harmonics is arbitrary. Clearly, one would like to pick those functions which simultaneously maximize the distortion while minimizing the noise. After experimenting with several different choices of weight functions we decided to use four Gaussian windows centered at 38, 58, 78, and 98 $h^{-1}$Mpc each with a dispersion of 8 $h^{-1}$Mpc. For this choice of weight functions, the off-diagonal elements of the covariance matrix, $A_{ij}$, are $\lesssim 10\%$ of the largest diagonal elements; thus the four spatial windows give roughly independent estimates of the redshift distortion. We will use these windows in all the analysis (both $N$-body and actual data) discussed below.

**TABLE 1**
CDM MONTE CARLO RESULTS

| Realization | $\beta$ |
|---|---|
| 1 | 0.93 |
| 2 | 1.11 |
| 3 | 1.06 |
| 4 | 0.86 |
| 5 | 0.83 |
| 6 | 0.88 |
| 7 | 1.15 |
| 8 | 1.13 |
| 9 | 0.51 |
| Mean: 0.94 | |
| Standard Deviation: 0.20 | |

## 4  Monte-Carlo Tests

In order to check the validity of our formalism, we applied the likelihood formalism to the real and redshift space weighted harmonics computed from an $N$-body simulation of a standard Cold Dark Matter universe characterized by $\Omega_o h = 0.5$. The simulations evolved $64^3$ particles in a box of 180 $h^{-1}$Mpc using the $P^3M$ algorithm until $rms$ variance of the density field in a sphere of 8 $h^{-1}$Mpc reached $\sigma_8 = 0.61$; further details of the simulations can be found in Frenk *et al.* [12]. In an effort to mimic current observational data, we extracted nine mock galaxy catalogues designed to closely match the properties of the 1.2 Jy *IRAS* redshift survey (cf., [25], [8]). Galaxy candidates were chosen as unbiased tracers of the underlying particle distribution and therefore $\beta = 1$ in the mock catalogues. The procedure for extracting these catalogues is described in Górski *et al.* [13] and Fisher *et al.* [9].

The $N$-body simulations provide a convenient way to test the formalism outline in the previous sections since both the amplitude and shape of the power spectrum are known *a priori*. Thus a simple test is to compute the weighted redshift harmonics for each of the mock *IRAS* catalogs and perform the likelihood analysis for the best value of $\beta$ holding the shape and amplitude of the power spectrum fixed at their correct values. Since our analysis is valid only in the linear regime, we restricted the likelihood computation to $l \leq 10$. Table 1 shows the resulting estimates of $\beta$ for the nine mock catalogs. The mean value of $\beta$ for the nine mock samples is $0.94 \pm 0.2$ which is in excellent agreement with the actual value $\beta = 1$. Thus, when the correct power spectrum is used the likelihood estimate for $\beta$ is unbiased; moreover the formal uncertainty in a single realization (given by $\Delta \mathcal{L} = -1/2$) is comparable to the scatter over the nine realizations, i.e., $\Delta \beta = 0.2$.

## 5  Application to the 1.2 Jy *IRAS* Survey

We now proceed to apply the formalism described in the previous section to redshift survey of 5313 *IRAS* galaxies flux-limited to 1.2 Jy at $60 \mu$m, selected from the *IRAS* database ([24], [8]). The sample covers 87.6% of the sky and is complete for $|b| > 5°$ with the exception of a small area of the sky not surveyed by *IRAS*. The SHA analysis discussed above relies on complete $4\pi$ steradian sky coverage. Although statistical corrections can be applied to the harmonics with partial sky coverage ([22], [23]), we have adopted a simpler method of dealing with incomplete sky coverage. We have interpolated the redshift data through the plane in a way which smoothly continues structure; this is

## TABLE 2
### PARAMETER CORRELATION MATRIX

|          | $\sigma_8$ | $\Gamma$ | $\beta$ |
|----------|------------|----------|---------|
| $\sigma_8$ | 1.00       | -0.16    | -0.81   |
| $\Gamma$   | -0.16      | 1.00     | 0.27    |
| $\beta$    | -0.81      | 0.27     | 1.00    |

the same approach used by Yahil *et al.* [27] in their reconstruction of the *IRAS* peculiar velocity field. Moreover, regularized mask inversion of the harmonics shows the effect of incomplete sky coverage is small for the *IRAS* 1.2 Jy sample geometry and $l \leq 10$ (Lahav *et al.* 1993, in preparation). The selection function, $\phi(r)$, is computed using the techniques of Yahil *et al.* [27]. In practice the selection function is computed by assigning luminosities based on galaxy redshifts, not distances; fortunately, the shape of the $\phi(r)$ is insensitive to the flow model used to correct for the effects of peculiar velocities [21]. Moreover, Monte-Carlo simulations have shown the results to be unaffected by the expected measurement redshift errors in the survey [11].

Since the distortion term in the harmonics given in Equation (9) is proportional to both the amplitude of the power spectrum ($\sigma_8^2$) and $\beta$, we expect a strong covariance in the likelihood estimates of these two parameters. Fortunately, the normalization of the real space power spectrum is fairly well established from analyses of the real space correlation function [10] and projected spherical harmonics [22]. As a first attempt, we computed the maximum likelihood estimates of $\beta$ and $\Gamma$ (the shape parameter of $P_R(k)$) holding the normalization of $P_R(k)$ fixed at the value determined by Fisher *et al.* [10], $\sigma_8 = 0.69 \pm 0.04$. The derived maximum likelihood values in the case are $\beta = 0.94 \pm 0.17$ and $\Gamma = 0.17 \pm 0.05$ and the corresponding likelihood contours are shown in Figure 1.

Allowing $\sigma_8$ to be a free parameter increases the statistical uncertainty in $\beta$ and leads to the following maximum likelihood values: $\beta = 0.47 \pm 0.25$, $\Gamma = 0.15 \pm 0.05$, and $\sigma_8 = 0.81 \pm 0.06$. Table 2 lists the correlation coefficients between the parameters. Table 2 shows the very strong anti-correlation between $\sigma_8$ and $\beta$. The shape parameter $\Gamma$, unlike $\sigma_8$ is positively correlated with $\beta$. Since the overall amplitude of the distortion is fixed by the data, an increase the predicted amount of large scale power (decreasing $\Gamma$) results in a lower inferred value of $\beta$. The recovered value of $\sigma_8$ when it is treated as a free parameter is just within the $3-\sigma$ error limits of the $\sigma_8$ determined from the correlation function although its statistical error $\sigma_8$ is 50% larger. Given the strong degree of $\beta$-$\sigma_8$ covariance, the more reliable value of $\beta$ is probably the one derived by treating $\sigma_8$ as a fixed parameter set by the real space correlation function ($\sigma_8 = 0.69$).

## 6  Discussion

The value of $\beta$ derived from the 1.2 Jy sample using the SHA technique is dominated by uncertainties in the amplitude of the power spectrum. The value of $\beta = 0.94 \pm 0.17$ determined by holding $\sigma_8$ fixed at 0.69, is in agreement with previous estimates from *IRAS* redshift surveys. Dekel *et al.* [4] found $\beta = 1.28^{+0.75}_{-0.59}$ ($2\sigma$) by comparing the density field of the 1.936 Jy *IRAS* survey [25] with the density field inferred from direct measurements of the peculiar velocity field using the POTENT algorithm ([1], [3]). Kaiser *et al.* [16] found $\beta = 0.9^{+0.20}_{-0.15}$ ($1\sigma$) in a comparison of the predicted *IRAS* peculiar velocity field with direct measurements using the deeper, albeit sparser, QDOT redshift survey. J. Roth (these proceedings) finds $\beta \sim 0.85 \pm 0.30$ based a comparison of directly measured peculiar velocities with the reconstructed maps of the *IRAS* 1.2 Jy peculiar velocity field (derived using the analysis of Yahil *et al.* [27]). Measurements based on the predicted acceleration of the Local Group using *IRAS* catalogs have also yielded high estimates $0.4 < \beta < 1.0$ ([20], [26]). Hamilton's [14] analysis of redshift distortions in the correlation function for the 1.936 Jy *IRAS* survey gave a slightly lower value, $\beta = 0.66^{+0.34}_{-0.22}$ ($1\sigma$), in good agreement with the value we obtained ($\beta = 0.47 \pm 0.25$) by allowing $\sigma_8$ to be a free parameter.

These results provide evidence of a high, near closure, value of $\Omega_\circ$ on large scales *if* the bias factor for *IRAS* is near unity. However, one should be cautious in claiming that these results provide definitive proof of $\Omega_\circ = 1$. Similar estimates of $\beta$ using optical surveys have yielded lower values ($0.25 < \beta < 0.65$) for the optical dipole [19] and ($\beta = 0.37 \pm 0.2$, $\beta = 0.5 \pm 0.06$) in comparisons of optical samples with peculiar velocities (E. Shaya and M. Hudson respectively, these proceedings)) suggesting that at least the ratio of the *IRAS* and optical bias factors may differ from unity.

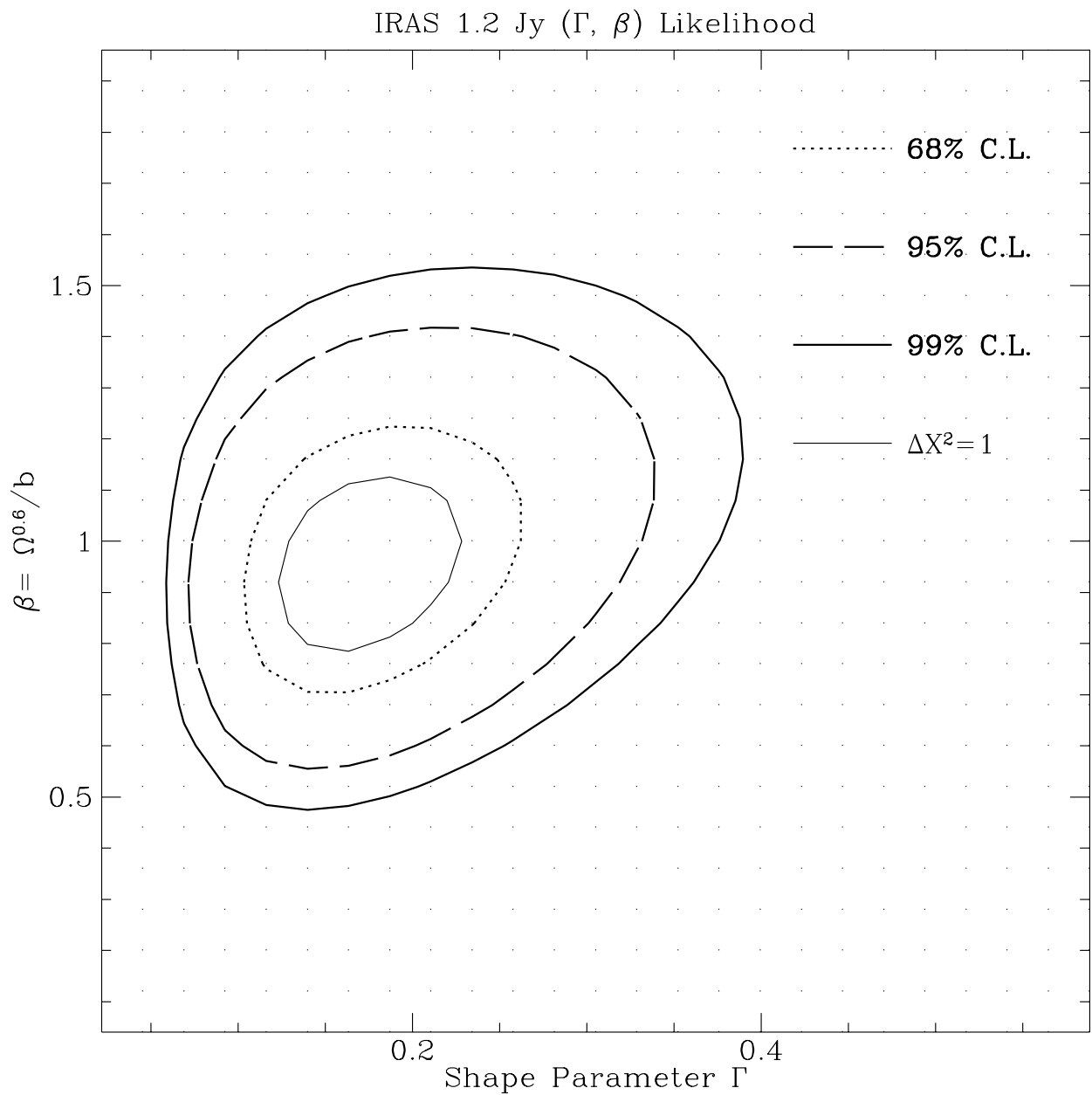

Figure 1: Likelihood contours derived from the SHA of the 1.2 Jy *IRAS* redshift survey. In this plot, the normalization of the power spectrum ($\sigma_8$) has been held fixed at 0.69 (the value determined independently from the real space correlation function). The distortion is therefore a sole function of $\beta$ and the shape of the power spectrum (parametrized by $\Gamma$). The contour levels correspond to the appropriate $\Delta\chi^2$ values for the confidence levels labeled in the figure for a likelihood function with two degrees of freedom.

Although the estimate of $\beta$ depends quite sensitively on the normalization of the power spectrum, the method gives a surprisingly robust measurement of the *shape* of the power spectrum with $\Gamma \simeq 0.16 \pm 0.05$. This result is somewhat lower that the value quoted by Feldman, Kaiser, & Peacock [7] of $\Gamma = 0.31 \pm 0.08$ in an analysis of the QDOT power spectrum and that of Mo, Peacock and Xia [18] of $\Gamma = 0.32 \pm 0.07$ deduced from the cluster-galaxy cross-correlation function. These measurements all argue strongly for a real space power spectrum that is significantly steeper on scales $\gtrsim 30$ $h^{-1}$Mpc than the canonical CDM model of $\Gamma = 0.5$ (ruled out in all three analyses at $\gtrsim$ 3-$\sigma$ level!). In the usual parlance, this can be phrased as the standard CDM model having an insufficient ratio of large to small scale power. Models containing a combination of hot and cold dark matter may provide one acceptable revision of the CDM model (cf. Pogoysan, these proceedings and references therein) as may the introduction of a cosmological constant [6].

The SHA analysis of redshift distortion presented here is well suited to near full-sky catalogs. The radial dependence of the selection function in flux-limited catalogs leads to a very compact expression for the distortion in terms of the harmonic power spectrum. The technique outlined in this paper, however, rests on the choice of an unspecfied radial weighting function, which induces the distortion. At first glance this arbitrariness may seem unappealing, however, it does allow the flexibility to optimize scheme for maximum distortion on a given scale. Alternatively, one can follow the approach used here and simply use a set of multiple weighting schemes which probe, albeit not optimally, a variety of different scales. Perhaps a more natural way to formulate the distortion is in terms of an expansion of the redshift space density field in spherical harmonics *and* orthogonal radial functions (see O. Lahav contribution to these proceedings). Spherical coordinates have the advantage that the redshift distortion couples only the radial modes thereby offering the possibility of removing the distortion using regularized inversion techniques. This work is currently in preparation and will be presented at a later date.

**Acknowledgements.** I would like to thank my collaborators Ofer Lahav and Caleb Scharf as well as Michael Strauss, John Huchra, Marc Davis, and Amos Yahil for permission to use the 1.2 Jy *IRAS* survey data prior to its publication.